\documentclass{iaus}
\usepackage{graphicx}

\title[IAU287.~~Radio and IR interferometry of SiO maser stars] %% give here short title %%
{Radio and IR interferometry\\ of SiO maser stars}

\author[Markus Wittkowski et al.]   %% give here short author list %%
{Markus Wittkowski$^1$,
David A. Boboltz$^2$,
Malcolm D. Gray$^3$, 
Elizabeth M. L. Humphreys$^1$
Iva Karovicova$^4$, \and
Michael Scholz$^{5,6}$}
\affiliation{
$^1$ESO, Karl-Schwarzschild-Str. 2, 85748 Garching bei M\"unchen, Germany\\[\affilskip]
$^2$US Naval Observatory, 3450 Massachusetts Avenue, NW, Washington, DC 20392-5420, USA\\[\affilskip]
$^3$Jodrell Bank Centre for Astrophysics, Alan Turing Building, University of Manchester, Manchester M13 9PL, UK\\[\affilskip]
$^4$Max-Planck-Institut f\"ur Astronomie, K\"onigstuhl 17, 69117 Heidelberg, Germany\\[\affilskip]
$^5$Zentrum f\"ur Astronomie der Universit\"at Heidelberg (ZAH), Institut f\"ur Theoretische Astrophysik, Albert-Ueberle-Str. 2, 69120 Heidelberg, Germany\\[\affilskip]
$^6$Sydney Institute for Astronomy, School of Physics, University of Sydney, Sydney NSW 2006, Australia
}

\pubyear{2012}
\volume{287}  %% insert here IAU Symposium No.
\pagerange{1--1}
\jname{Cosmic masers - from OH to H$_0$}
\editors{R.S. Booth, E.M.L. Humphreys, W.H.T. Vlemmings, eds.}
\begin{document}

\maketitle

\begin{abstract}
Radio and infrared interferometry of SiO maser stars provide
complementary information on the atmosphere and
circumstellar environment at comparable spatial resolution.
Here, we present the latest results on the
atmospheric structure and the dust condensation region
of AGB stars based on our recent infrared spectro-interferometric observations,
which represent the environment of SiO masers.
We discuss, as an example, new results from simultaneous
VLTI and VLBA observations of the Mira variable AGB star R~Cnc, including
VLTI near- and mid-infrared interferometry, as well as VLBA observations
of the SiO maser emission toward this source. We present preliminary
results from a monitoring campaign of high-frequency SiO maser emission
toward evolved stars obtained with the APEX telescope, which also serves
as a precursor of ALMA images of the SiO emitting region.
We speculate that large-scale long-period chaotic motion in the extended
molecular atmosphere may be the physical reason for observed deviations
from point symmetry of atmospheric molecular layers, and for
the observed erratic variability of high-frequency SiO maser emission.

\end{abstract}

\firstsection % if your document starts with a section,
\section{Introduction}
Low to intermediate mass stars, including our Sun, evolve to red giant
and subsequently to asymptotic giant branch (AGB) stars.
An AGB star is in the final stage of stellar evolution that is driven by
nuclear fusion.
Mass loss becomes increasingly important during
the AGB evolution, both for the stellar evolution, and for the return of
material to the interstellar medium.
Depending on whether or not carbon has been dredged up from the core into
the atmosphere,
AGB stars appear to have an oxygen-rich or a carbon-rich
chemistry. A canonical model of the mass-loss process has been developed for
the case of the carbon-rich chemistry, where atmospheric carbon dust has a
sufficiently large opacity to be radiatively accelerated and driven out of
the gravitational potential of the star and where it drags along the gas
(e.g., \cite[Wachter et al. 2002]{wachter02}, \cite[Mattsson et al. 2010]{mattson10}).
For the case of an oxygen-rich chemistry, the details of this process are
not understood, and are currently a matter of debate 
(e.g., \cite[Woitke 2006]{woitke06}, \cite[H\"ofner 2008]{hoefner08}).
Questions remain also
for the carbon-rich case, such as regarding the recent observational
evidence that the oxygen-bearing molecule
H$_2$O is ubiquitous in C-stars (\cite[Neufeld et al. 2011]{neufeld11}).
Another unsolved problem in stellar physics is the mechanism by which
(almost) spherically symmetric stars on the AGB
evolve to form axisymmetric planetary nebulae (PNe). Currently, a consensus
seems to form that single stars cannot trivially evolve towards
non-spherical PNe, by mechanisms such as magnetic fields or rotation, but
that a binary companion is required in the majority of the observed
PN morphologies (e.g. \cite[de Marco 2009]{demarco09}).

In order to solve these open questions, it is important to 
observationally establish the detailed stratification and geometry of the 
extended atmosphere and the dust formation region, and to compare it to and 
constrain the different modeling attempts. This includes the following
questions: How is the mass-loss process connected to the stellar pulsation?
Which is the detailed radial structure of the atmosphere and circumstellar
envelope? At which layer do inhomogeneities form? Which are the shaping
mechanisms?  Which is the effect of inhomogeneities on the further stellar
evolution?

\section{Synergy of radio and infrared interferometry}
Radio and infrared interferometry are both well suited to probe the
structure, morphology, and kinematics of the extended atmospheres
and circumstellar environment of evolved stars, because of their
ability to spatially resolve these regions.
In fact, evolved stars have been prime targets for radio and infrared 
interferometry for decades, because they match well the 
sensitivity and angular scale of available facilities.
Interferometric observations continue to provide new observational
results in the field of evolved stars thanks to newly available
spectro-interferometry at infrared wavelengths, new radio
interferometric facilities, larger samples of well studied targets, and 
comparisons to newly available theoretical models. 
Here, radio and infrared interferometry provide complementary information.
From the perspective of infrared studies, radio interferometry of 
SiO, OH, and H$_2$O maser emission adds information on the morphology
and kinematics at different scales from a few stellar radii (SiO maser)
to a few hundred stellar radii (OH maser). From the perspective
of radio interferometry, infrared interferometry provides information
on the environment of the astrophysical masers, including the radiation
field, the radii at wavelengths of maser pumping, and constraints on the
stratification of the temperature, density, and number densities.

\section{Project outline}
We have established a project of coordinated
interferometric observations of evolved stars at infrared and radio wavelengths.
Our goal is to constrain 
the radial structure and kinematics of the stellar atmosphere and the
circumstellar environment to understand better the mass-loss process
and its connection to stellar pulsation.
We also aim at tracing asymmetric structures from small to large
distances in order to constrain shaping processes during the
AGB evolution.
We use two of the highest resolution interferometers in the world,
the Very Large Telescope Interferometer (VLTI) and the Very Long Baseline
Array (VLBA) to study AGB stars and their circumstellar envelopes
from near-infrared to radio wavelengths. For some sources, we have
coordinated near-infrared broad-band photometry obtained at the
South African Astronomical Observatory (SAAO) in order to derive
effective temperatures.
We have started to
use the Atacama Pathfinder Experiment (APEX) to investigate the line strengths
and variability of high frequency SiO maser emission in preparation of 
interferometric observation of SiO emitting regions using the ALMA facility.

\section{Observations}
Our pilot study included coordinated observations of the Mira variable
S Orionis including VINCI $K$-band measurements at the VLTI and SiO maser
measurements at the VLBA (\cite[Boboltz \& Wittkowski 2005]{boboltz05}). For the Mira
variables S Ori, GX Mon, RR Aql and the supergiant AH Sco, we obtained
long-term mid-infrared interferometry covering several pulsation cycles
using the MIDI instrument at the VLTI
coordinated with VLBA SiO (42.9 GHz and 43.1 GHz transitions) observations
(partly available in \cite[Wittkowski et al. 2007]{wittkowski07},
\cite[Karovicova et al. 2011]{karovicova11}).
For the Mira variables R~Cnc and X~Hya, we
coordinated near-infrared interferometry (VLTI/AMBER),
mid-infrared interferometry (VLTI/MIDI), VLBA/SiO maser observations,
VLBA/H$_2$O maser, and near-infrared photometry at the SAAO 
(first results in \cite[Wittkowski et al. 2008]{wittkowski08}).
Most recently, we obtained measurements of the $v=1$ and $v=2$ $J=7-6$ SiO maser
transitions toward a sample of evolved stars at several epochs
using the APEX telescope.

\section{Modeling}
\label{sec:modeling}
We used the P \& M model series by \cite[Ireland et al. (2004a,b)]{ireland04a,ireland04b}
and most recently the CODEX models by \cite[Ireland et al. (2008, 2011)]{ireland08,ireland11}
to describe the dynamic model atmospheres of Mira variable
AGB stars. 
\cite[Wittkowski et al. (2007)]{wittkowski07} and \cite[Karovicova et al. (2011)]{karovicova11}
added ad-hoc radiative transfer models to these dynamic model 
atmosphere series to describe the dust shell. They employed 
the radiative transfer code {\tt mcsim\_mpi} by \cite[Ohnaka et al. (2007)]{ohnaka07}.
\cite[Gray et al. (2009)]{gray09} combined
these hydrodynamic atmosphere plus dust shell models with a maser
propagation code in order to model the SiO maser emission.

\cite[Wittkowski et al. (2007)]{wittkowski07} conducted coordinated mid-infrared 
interferometry using the VLTI/MIDI instrument and 
observations of SiO maser emission using the VLBA of the Mira
variable S Ori. Based
on the modeling of the mid-infrared interferometry as outlined
above, they showed that the maser emission is located just outside
the layer where the molecular layer becomes optically thick
at mid-infrared wavelengths ($\sim 10\mu$m), roughly at two
photospheric radii, and is close to, and possibly co-located 
with Al$_2$O$_3$ dust. 
One of the questions that led to the modeling effort by
\cite[Gray et al. (2009)]{gray09} was whether the combined dynamic model atmosphere
and dust shell model that was successful to describe the 
mid-infrared interferometric observations of S Orionis by 
\cite[Wittkowski et al. (2007)]{wittkowski07} would also lead to model-predicted 
locations of SiO maser emission that is consistent with the
simultaneous VLBA observations. The input to the maser propagation
model included the temperature and density stratification of the
successful model of the mid-infrared interferometric data,
the radii of the 1.04\,$\mu$m continuum layer, and of optically
thick layers at IR pumping bands of SiO (8.13\,$\mu$,
4.96\,$\mu$, 2.71\,$\mu$, and 2.03\,$\mu$m), the IR radiation 
field of the dust model, and the number densities of SiO and its
main collision partners (assuming LTE chemistry). 
Modeled masers indeed formed in rings with
radii of 1.8--2.4 photospheric radii, which is consistent
with the S Ori VLBA observations, with other observations in the 
literature, and with earlier such maser propagation models
(\cite[Gray et al. 1995]{gray95}, \cite[Humphreys et al. 1996]{humphreys96}). The new models confirm the $v=1$
ring at larger radii than the $v=2$ ring. Maser rings, a shock front,
and the 8.13\,$\mu$m layer appear to be closely related, suggesting
that collisional and radiative pumping are closely related spatially
and therefore temporally.

\section{Infrared and radio interferometry of the Mira variable AGB 
star R~Cnc}

R Cancri (R~Cnc) is a Mira variable AGB star with a V magnitude
between 6.1 and 11.8 and a period of 362 days (\cite[Samus et al. 2009]{samus09}) 
at a distance of 280 pc based on the period-luminosity relation by
\cite[Whitelock et al. (2008)]{whitelock08}. We obtained two epochs of observations
(23 Dec 2008 to 10 Jan 2009 and 25 Feb 2009 to 3 Mar 2009) that
were coordinated between near-infrared spectro-interferometry
obtained with VLTI/AMBER, mid-infrared spectro-interferometry
obtained with VLTI/MIDI, near-IR $JHKL$ photometry obtained
at the SAAO, and VLBA observations of the SiO maser emission
and H$_2$O maser emission. The details of these observations
will be available in Wittkowski et al. (in preparation).
The first epoch of near-infrared spectro-interferometry of R~Cnc is
available in \cite[Wittkowski et al. (2011)]{wittkowski11}.

\subsection{Near-infrared spectro-interferometry of R~Cnc}
\begin{figure}[b]
\begin{center}
\includegraphics[width=10cm]{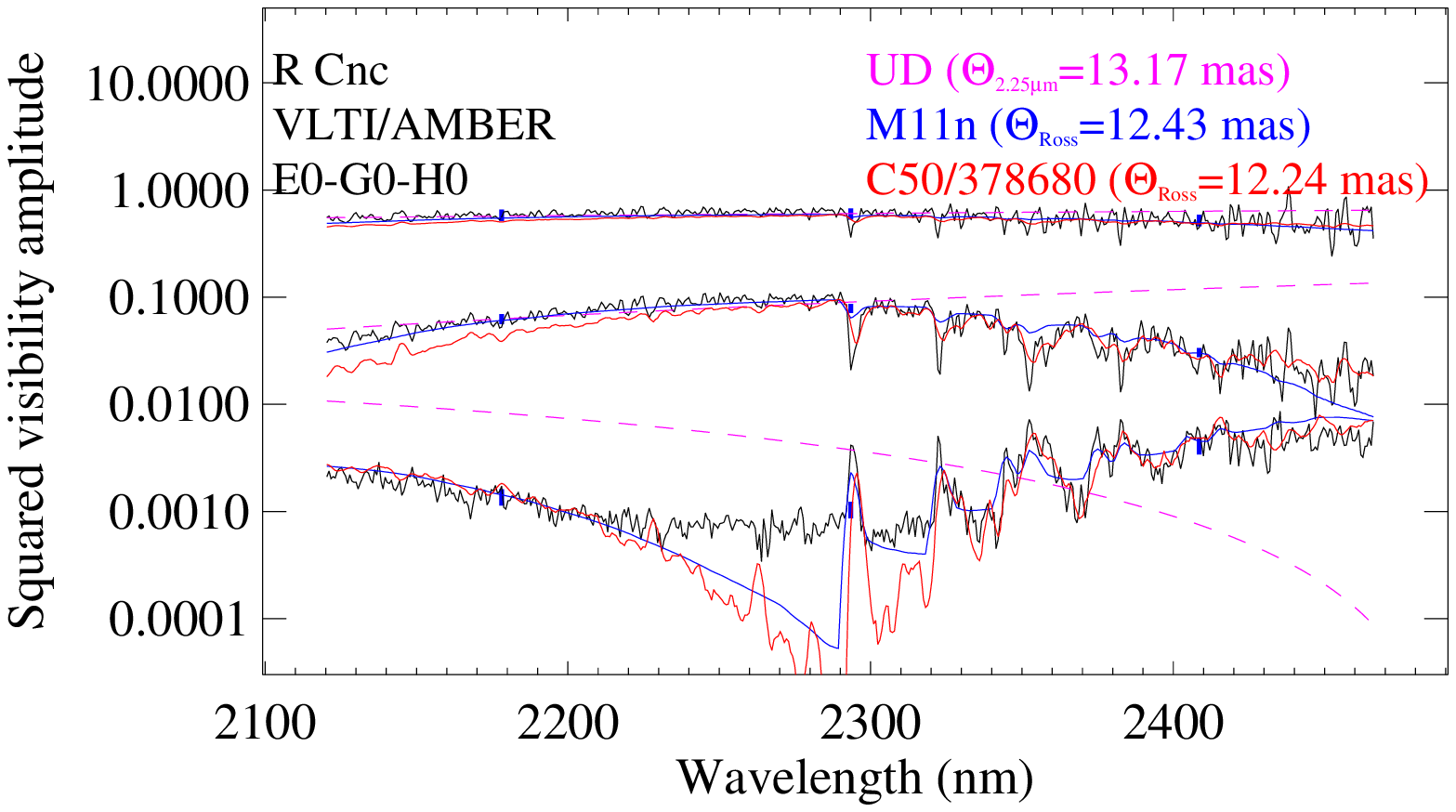}
\includegraphics[width=10cm]{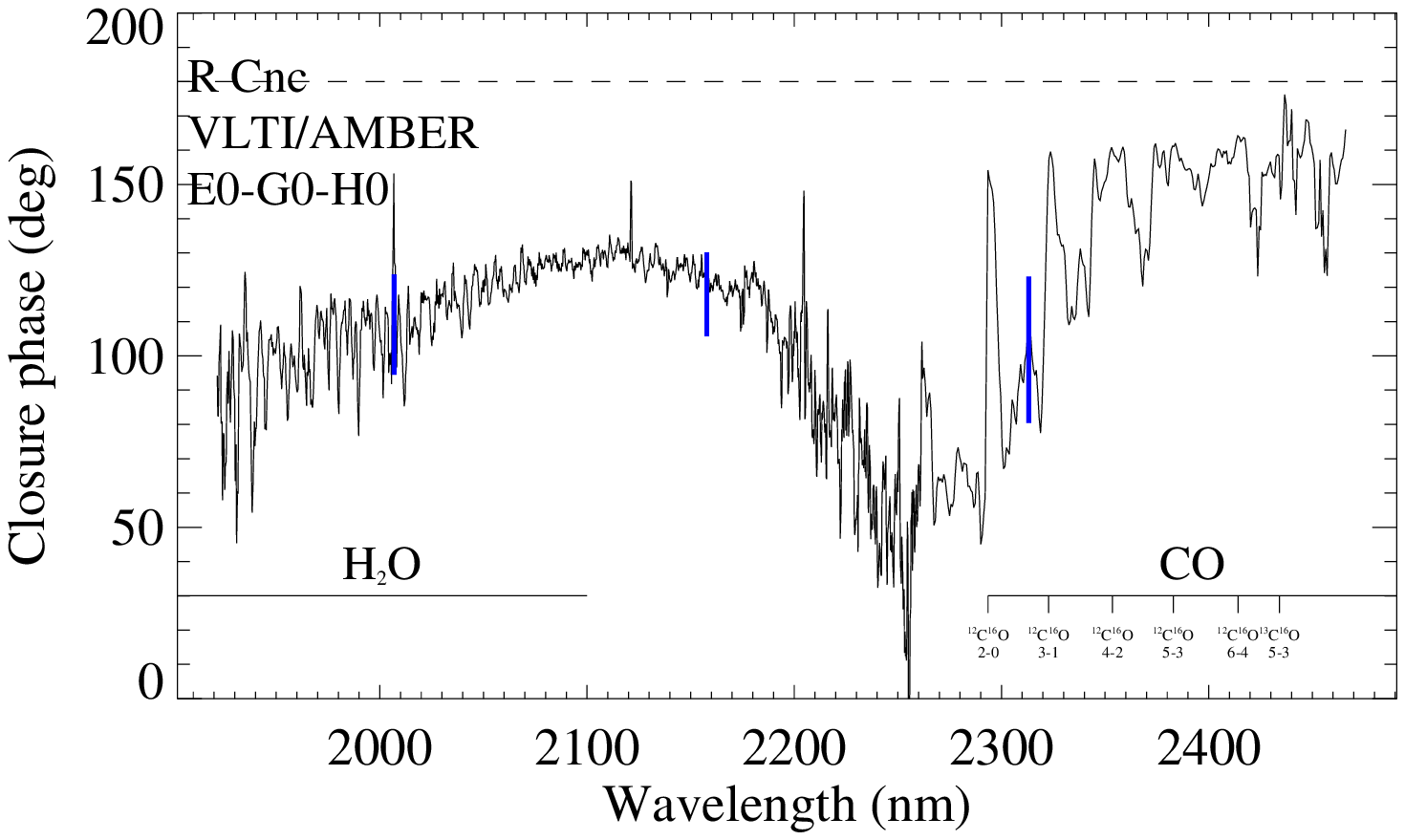}
\caption{Near-IR squared visibility amplitude (top)
and closure phase (bottom) of R~Cnc obtained
with the VLTI/AMBER instrument from \cite[Wittkowski et al. (2011)]{wittkowski11}. The 
upper panel also shows predictions by dynamic model
atmospheres (M series by \cite[Ireland et al. 2004a,b]{ireland04a,ireland04b},
and CODEX series by \cite[Ireland et al. 2008, 2011]{ireland08,ireland11}).}
\label{fig:rcncamber}
\end{center}
\end{figure}
Fig.~\ref{fig:rcncamber} shows the near-infrared 
squared visibility amplitudes and closure phases of R~Cnc
obtained with the VLTI/AMBER instrument (from \cite[Wittkowski et al. 2011]{wittkowski11}).
The visibility amplitudes show a characteristic {\it bumpy} shape
that has also been observed in VLTI/AMBER observations of 
S Ori in \cite[Wittkowski et al. (2008)]{wittkowski08} and that is interpreted as being
indicative of the presence of molecular layers lying on top
of the continuum-forming photosphere and that extend to a
few photospheric radii. The visibilities are well consistent 
with predictions by the latest dynamic model atmosphere series
by \cite[Ireland et al. (2008, 2011)]{ireland08,ireland11} that include such
molecular layers. The wavelength-dependent closure phases 
indicate deviations from point symmetry at all wavelengths
and thus a complex non-spherical stratification of the 
atmosphere. In particular, \cite[Wittkowski et al. (2008)]{wittkowski08}
discuss that the strong closure phase signal
in the water vapor and CO bandpasses can be a 
signature of large-scale inhomogeneities/clumps of the 
molecular layers. These might be caused by pulsation- and
shock-induced chaotic motion in the extended 
atmosphere as theoretically predicted by \cite[Icke et al. (1992)]{icke92}
and \cite[Ireland et al. (2008, 2011)]{ireland08,ireland11}. We note that 
these extended atmospheric layers correspond roughly to the
radii where SiO maser emission is observed.

\subsection{Mid-infrared spectro-interferometry of R~Cnc}
\begin{figure}[b]
\begin{center}
\includegraphics[width=6cm,angle=90]{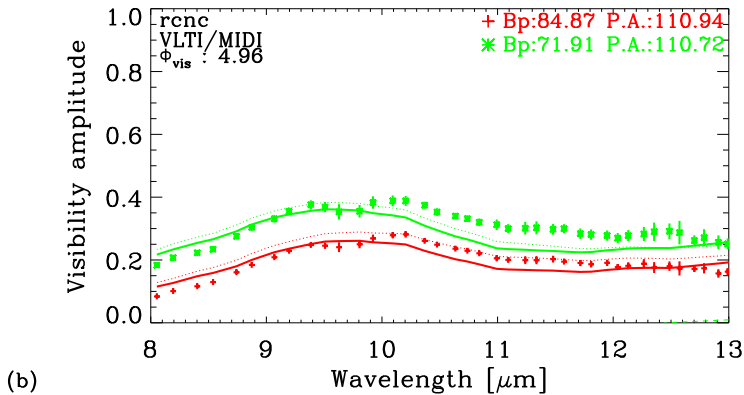}
\caption{Mid-IR visibility amplitude
of R~Cnc obtained
with VLTI/MIDI from \cite[Karovicova (2011)]{karovicovaphd}. Also
shown is a model of a dynamic model atmosphere
combined with a radiative transfer model of the
dust shell (see text).}
\label{fig:rcncmidi}
\end{center}
\end{figure}

Fig.~\ref{fig:rcncmidi} shows the 
mid-IR visibility amplitude of R~Cnc obtained
with VLTI/MIDI from \cite[Karovicova (2011)]{karovicovaphd}.
The MIDI data can well be re-produced with a 
dynamic model atmosphere, which naturally includes
molecular layers, and an ad-hoc radiative transfer
model of an Al$_2$O$_3$ dust
shell with an inner radius of $\sim 2.2$ photospheric
radii and an optical depth $\tau_V\sim 1.4$.
As in the case of MIDI observations of S Ori 
(\cite[Wittkowski et al. 2007]{wittkowski07}), R~Cnc does not show an indication
of an additional silicate dust shell. The
inner radius of the Al$_2$O$_3$ dust
shell is located at a radius that is close to the radius
where SiO maser emission is observed.
Oher sources that have been studied
by \cite[Karovicova (2011)]{karovicovaphd} show either Al$_2$O$_3$ dust,
silicate dust, or both dust species. 
\cite[Karovicova (2011)]{karovicovaphd} discussed an indication that 
the dust content of stars with low mass-loss rates is
dominated by Al$_2$O$_3$, while the dust content of stars
with higher mass-loss rates predominantly exhibit
significant amounts of silicates, as suggested by
\cite[Little-Marenin \& Little (1990)]{little90} and 
\cite[Blommaert et al. (2006)]{blommaert06}.

\subsection{VLBA observation of the SiO maser
emission toward R~Cnc}
\begin{figure}[b]
\begin{center}
\includegraphics[width=7cm]{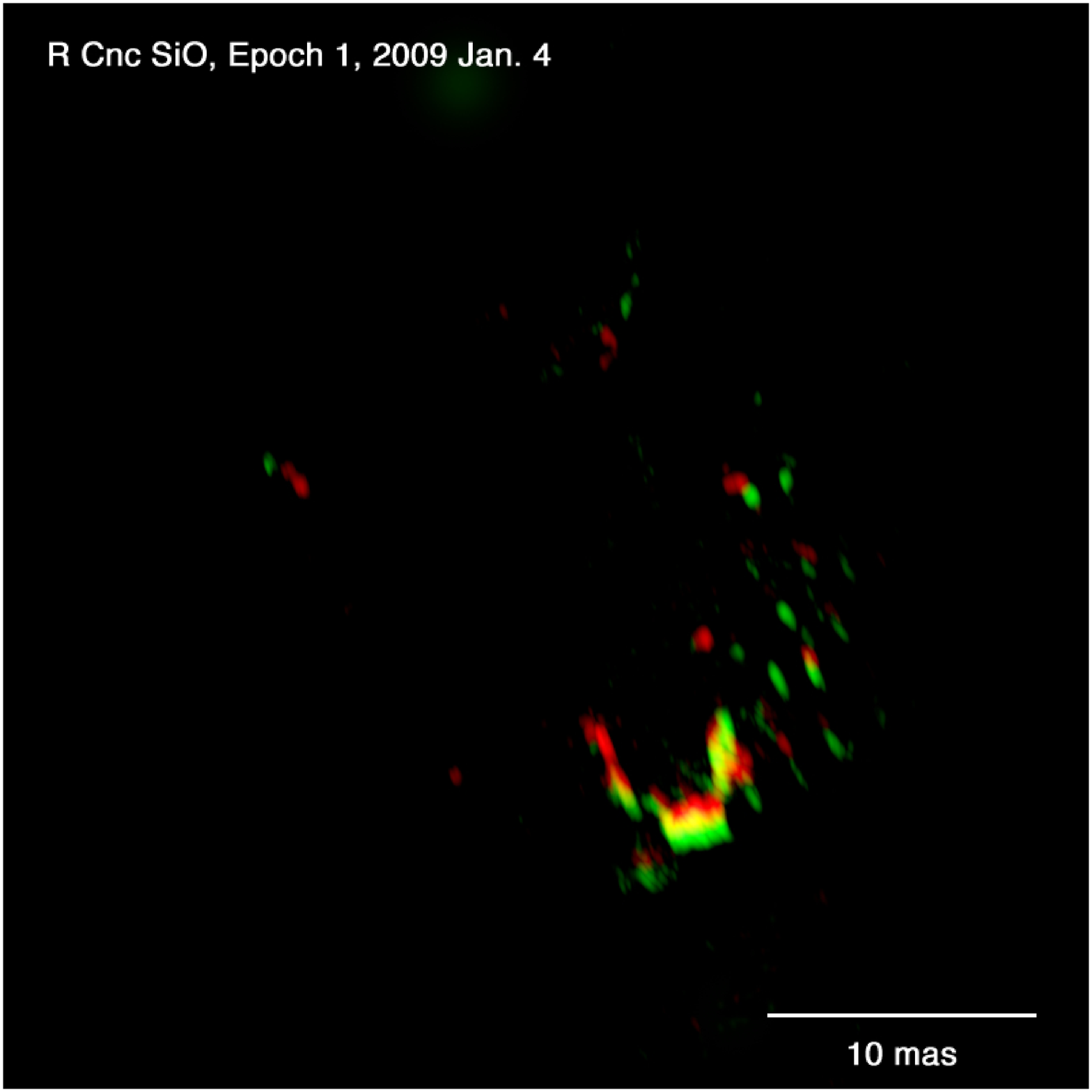}
\includegraphics[width=7cm]{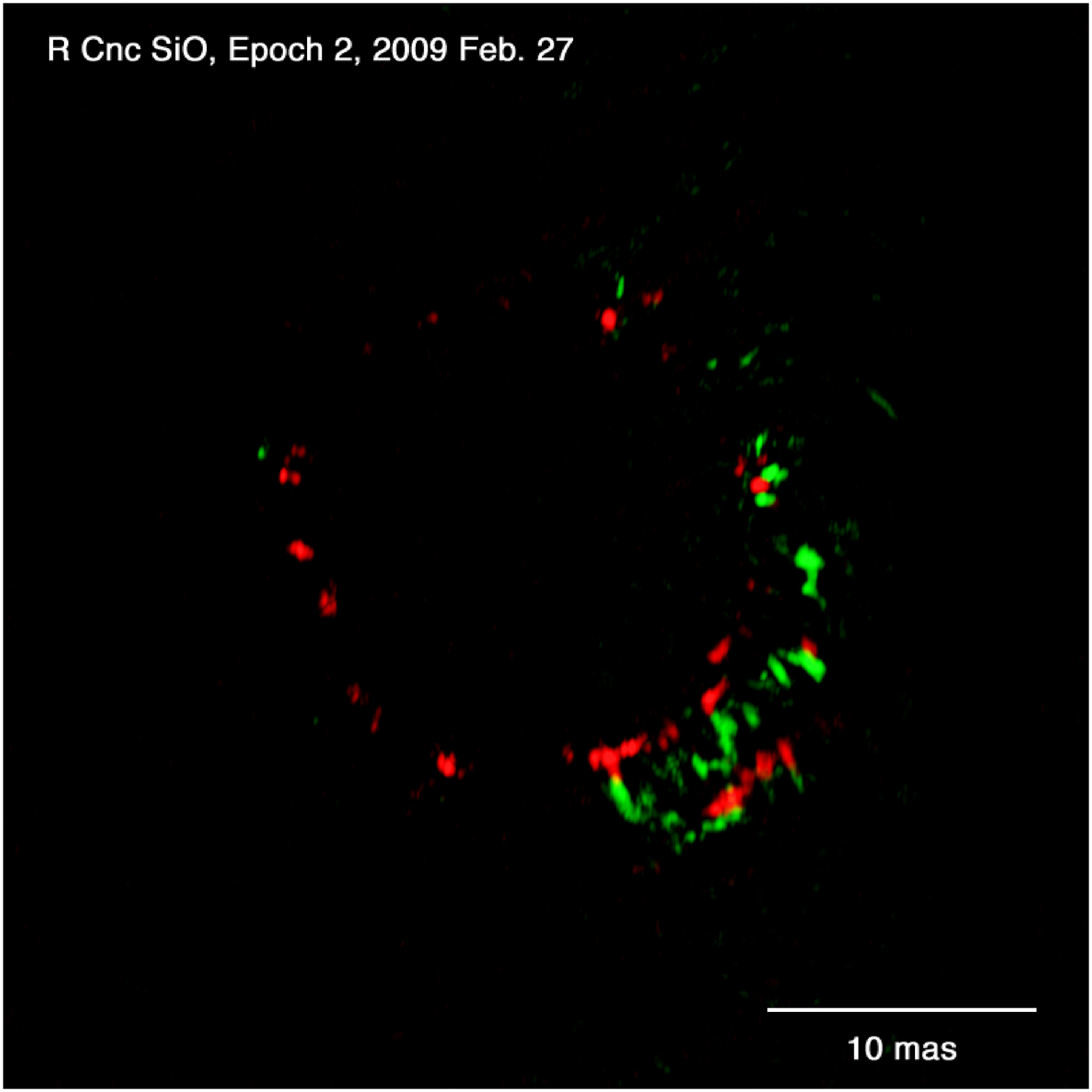}
\caption{VLBA observations of the SiO emission toward
R~Cnc at two epochs (top: 4 Jan 2009, bottom: 27 Feb 2009). 
The red maps denotes the $J=1-0$, $v=2$ transition
and the green maps the $J=1-0$, $v=1$ transition.}
\label{fig:rcncsio}
\end{center}
\end{figure}
Fig.~\ref{fig:rcncsio} shows our two epochs of 
VLBA images of the $J=1-0$, $v=2$ (42.8 GHz)
and $J=1-0$, $v=1$ (43.1 GHz) SiO maser emission
toward R~Cnc. The two epochs are separated by
about 7 weeks.
The two transitions were registered
to each other by transferring the calibration 
from one transition to the other. Consistently
with earlier observations, the $v=1$ transition is
located at larger radii than the $v=2$ transition,
but with more overlap at epoch 1 and a clearer
separation at epoch 2. The morphology is more
ring-like at epoch 2 compared to epoch 1.

\section{Monitoring of high-frequency SiO maser emission
toward evolved stars}
\begin{figure}[b]
\begin{center}
\includegraphics[width=14cm]{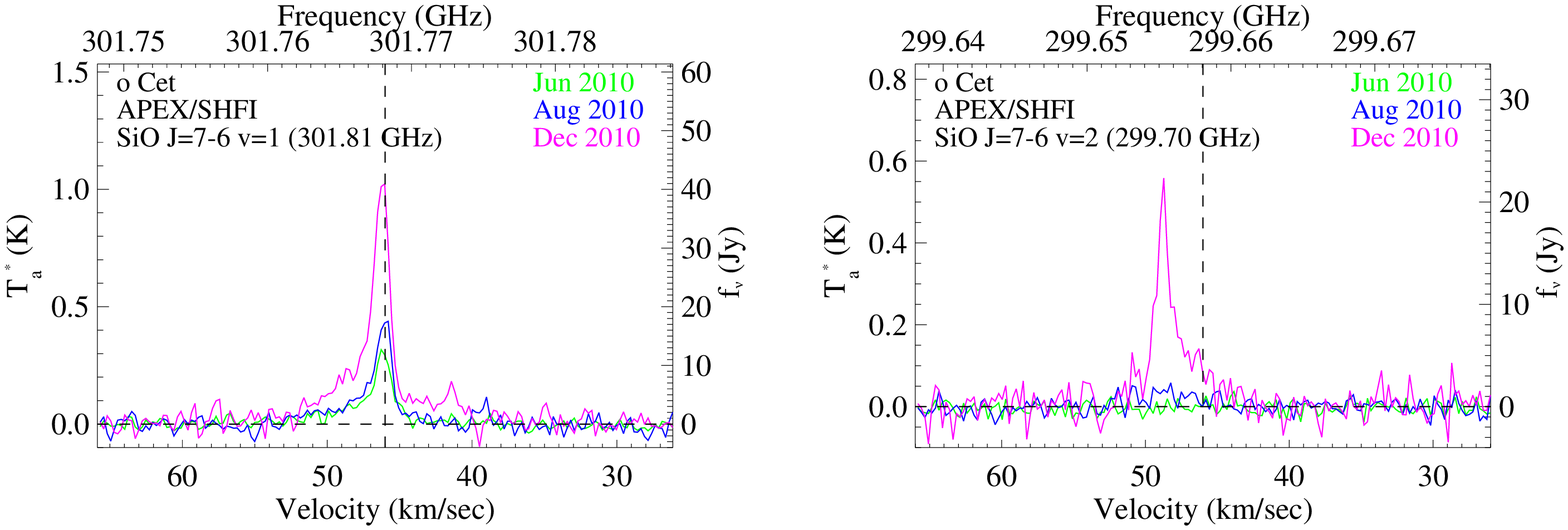}
\caption{APEX observations of the $J=7-6$,
$v=1$ (left) and $v=2$ (right) SiO maser emission
of o Cet at three epochs in June 2010 (phase 0.5),
August 2010 (phase 0.7), and December 2010 (phase 1.1).}
\label{fig:omicet}
\end{center}
\end{figure}

Observations of high-frequency SiO maser emission
has been shown to be more variable than centimeter 
SiO maser emission, indicating
that high-frequency maser emission depends more 
strongly on the environmental
conditions of the masers, and that thus their 
observations may provide stronger
observational constraints (\cite[Gray et al. 1995]{gray95}, 
\cite[Humphreys et al. 1997]{humphreys97}, \cite[Gray et al. 1999]{gray99}).

Following these studies, we have established a monitoring of high-frequency
$J=7-6$, $v=1,2,3$ SiO maser emission toward a sample of AGB stars in order
to compare their variability to the predictions by \cite[Gray et al. (2009)]{gray09}.
This represents also an important precursor of ALMA imaging studies of the 
SiO emitting regions of evolved stars.

Our observational results suggest that the variability is erratic. It does
not appear to be correlated with the stellar phase and is also not consistent
among the different sources of our sample. For, example, o Cet showed
$v=1$ emission at all phases between stellar minimum and post-maximum with
increasing intensity, and $v=2$ emission only at post-maximum
(phase 1.1) with lower intensity compared to $v=1$ emission at this phase
(Fig.~\ref{fig:omicet}).
To the contrary, R Hya, for instance, shows $v=1$ emission only at 
post-maximum (1.1) but not between minimum and maximum, 
but $v=2$ emission at all phases
between 0.5 and 1.1 and at phase 1.1 with higher intensity compared to $v=1$.
R~Leo showed both $v=1$ and $v=2$ emission at phases between 0.5 and 0.8
with alternating ratio between the strength of the $v=1$ and $v=2$ emission.

We hypothesize that large-scale (a few cells across the stellar
surface) long-period (times scales corresponding to a few 
pulsation cycles) chaotic motion
in the extended atmosphere, induced by the interaction of
pulsation and shock fronts with the extended atmosphere,
and a possibly related erratic variability of the SiO
abundance, may be the reason for our observed
erratic maser variability.

\section{Summary}
Near-infrared interferometry of oxygen-rich evolved stars 
indicates a complex atmosphere including extended
atmospheric molecular layers
(in the IR most importantly H$_2$O, CO, SiO), which is 
consistent with predictions
by the latest dynamic model atmospheres. Near-IR closure
phases indicate a complex non-spherical stratification of the
atmosphere, indicating asymmetric/clumpy molecular layers.
These are possibly caused by chaotic motion in the 
extended atmosphere, which may be triggered by the pulsation
in the stellar interior. Mid-infrared interferometry
constrains dust shell parameters including Al$_2$O$_3$
dust with inner radii of typically two photospheric radii
and silicate dust with inner radii of typically four photospheric
radii.
SiO masers lie in the extended atmosphere as seen by
infrared interferometry. They are located just outside
the radius where the molecular layer becomes optically
thick at mid-IR wavelengths. There are located close-to,
possibly co-located, with Al$_2$O$_3$
dust. Their location is consistent with
dynamic model atmospheres combined with a maser propagation
code. They are also located at the distances where the
near-IR interferometry indicates a clumpy morphology.
APEX detects a strong and erratic variability of high-frequency
maser emission. We speculate that the erratic variability may
be connected to chaotic motion in the extended atmosphere, i.e. 
to the same mechanism that may lead to the observed clumpyness
of extended atmospheric molecular layers.

\acknowledgements
This article is based on a project of coordinated VLTI and VLBA
observations of evolved stars to which several people have
contributed during the last years in addition to the authors of 
this article, including Carlos de Breuck,
Thomas Driebe, Eric Fossat, Michael Ireland,
Keiichi Ohnaka, Anita Richards,
Francois van Wyk, Patricia Whitelock, Peter Wood, and Albert Zijlstra.

{}

\end{document}